\documentclass[preprint,nopacs,nofootinbib]{revtex4}

\usepackage{amsfonts}
\usepackage{amsmath}
\usepackage{mathtools}
\usepackage{xcolor}
\usepackage{bm}
\usepackage{graphicx}
\usepackage{cancel}

\usepackage{hyperref}
\usepackage{cancel}

\usepackage{tipa}

\usepackage{wasysym}

\usepackage{tcolorbox}

\newcommand{\be}{\begin{equation}}
\newcommand{\ee}{\end{equation}}
\newcommand{\ba}{\begin{eqnarray}}
\newcommand{\ea}{\end{eqnarray}}

\newcommand{\bomega}{\boldsymbol{\omega}}

\newcommand{\bB}{\boldsymbol{B}}

\newcommand{\bM}{\boldsymbol{M}}

\newcommand{\bR}{\boldsymbol{R}}

\DeclareMathAlphabet{\mathpzc}{OT1}{pzc}{m}{it}

\newcommand{\bbe}{\boldsymbol{\mathrm{e}}}

\newcommand{\bdiff}{\textrm{\bf d}}
\newcommand{\bDiff}{\textrm{\bf D}}

\newcommand{\lp}{\left(}
\newcommand{\rp}{\right)}

\newcommand{\+}{ \prescript{+}{}}

\begin{document}

\title{The $\Lambda$ and the CDM as integration constants}

\author{Priidik Gallagher}
\email{priidik.gallagher@ut.ee}
\affiliation{Laboratory of Theoretical Physics, Institute of Physics,
University of Tartu, W. Ostwaldi 1, 50411 Tartu, Estonia}

\author{Tomi Koivisto}
\email{tomik@astro.uio.no}
\affiliation{Laboratory of Theoretical Physics, Institute of Physics,
University of Tartu, W. Ostwaldi 1, 50411 Tartu, Estonia} 
\affiliation{National Institute of Chemical Physics and Biophysics, R\"avala pst. 10, 10143 Tallinn, Estonia}
\affiliation{University of Helsinki and Helsinki Institute of Physics, P.O. Box 64, FI-00014, Helsinki, Finland}

\date{\today}

\begin{abstract}

Notoriously, the two main problems of the standard $\Lambda$CDM model of cosmology are the cosmological constant $\Lambda$ and the cold dark matter, CDM. 
This essay shows that both the $\Lambda$ and the CDM arise as integration constants in a careful derivation of Einstein's equations from first principles in a Lorentz gauge theory. 
The dark sector of the universe might only reflect the geometry of a spontaneous symmetry breaking that is necessary for the existence of a spacetime and an observer therein. 

\end{abstract}

\maketitle


\section{Introduction}

General relativity is the local version of special relativity. Gravity is thus understood to be a gauge theory of the Lorentz group. The basic variable is then a Lorentz connection 1-form $\bomega^a{}_b$, which defines the covariant derivative $\bDiff$, and thereby the curvature 2-form $\bR^a{}_b=\bdiff \bomega^a{}_b+\bomega^a{}_c\wedge\bomega^c{}_b$ subject to the 3-form Bianchi identity $\bDiff \bR^a{}_b= 0$ inherited from the Jacobi identity of the Lorentz algebra.

Since the beginning  \cite{Utiyama:1956sy}, the role of translations in the inhomogeneous Lorentz group has been elusive. What has been clear is that, in order to recover the dynamics of general relativity, some extra structure is required besides the connection $\bomega^a{}_b$. The standard approach since Kibble's work \cite{Kibble:1961ba} has been to introduce the coframe field $\bbe^a$, another 1-form valued in Lorentz algebra. Only recently, the more economical
possibility of introducing solely a scalar field $\tau^a$, was put forward by Z\l{}o\'snik {\it et al} \cite{Zlosnik:2018qvg}. Only then is gravity described by variables that are fully analogous to the fields of the 
standard Yang-Mills theory. 
 
The symmetry-breaking scalar $\tau^a$ has been called the (Cartan) {\it Khronon} because it encodes the foliation of spacetime. The theory of Z\l{}o\'snik {\it et al} is pre-geometric in the sense that there exist symmetric solutions (say $\tau^a=0$) where there is no spacetime. Only in a spontaneously broken phase $\tau^2<0$, there emerges a coframe field $\bbe^a = \bDiff\tau^a$ and further, if the coframe field is non-degenerate, a metric tensor $g_{\mu\nu}\bdiff x^\mu\otimes\bdiff x^\nu = \eta_{ab}\bbe^a\otimes\bbe^b$. In terms of the 2 fundamental fields, the Lorentz connection and the {\it Khronon} scalar, the theory realises the idea of observer space \cite{Gielen:2012fz}.  

A serendipitous discovery was that in the broken phase the theory does not quite reduce to general relativity, but to general relativity with dust  \cite{Zlosnik:2018qvg}. The presence of this ``dust of time'' could explain the
cosmological observations without dark matter. In this essay, we shall elucidate how this geometrical dark matter appears as an integration constant at the level of field equations. In addition,
we consider the next-to-simplest model by introducing the cosmological $\Lambda$-term. This will require another symmetry-breaking field, the (Weyl) {\it Kairon} $\sigma^a$, which turns out to impose unimodularity. 

The conclusion we wish to present is that a minimalistic gauge theory of gravity includes both the $\Lambda$ and CDM, and they both enter into the field equations as integration constants 
in the broken phase. 

\section{Dark matter}

Let us first make the case for dark matter. In the original, quite dense article  \cite{Zlosnik:2018qvg} the result was derived by a Hamiltonian analysis that may not be easy to follow in details. Therefore, we believe the simple derivation below could be useful. 

The SO(4,$\mathbb{C}$) action can be written in the quadratic form  \cite{Zlosnik:2018qvg},  
\be \label{action}
I_G = \frac{1}{2}\int \lp \frac{1}{8}\tau^2\epsilon^{bd}{}_{ac}+i\tau^b\tau^d\eta_{ac}\rp\bR^a{}_b\wedge\bR^c{}_d\,,
\ee
and the variations with respect to the two fields yield (what have been called ``the infernal equations''),
\begin{subequations}
\label{efe}
\ba
\bDiff(\prescript{+}{}\bR^a{}_b\wedge \bDiff \tau^b) & = & 0\,,  \label{efe1} \\
\frac{1}{2}\bDiff \prescript{+}{}(\bDiff\tau^{[a}\wedge\bDiff\tau^{b]}) & = & \tau^{[a}\+\bR^{b]}{}_c\wedge\bDiff\tau^c\,. \label{efe2}
\ea
\end{subequations}
The anti/self-dual projections of a field $X^a{}_b$ in the adjoint representation are denoted as $\prescript{\pm}{}X^a{}_b$ and defined by the property $\epsilon^{ad}{}_{bc}\prescript{\pm}{}X^c{}_d = \pm 2i\prescript{\pm}{}X^{a}{}_b$. There emerges a formal solution to (\ref{efe1}),
\be \label{solution}
\prescript{+}{}\bR^a{}_b\wedge \bDiff \tau^b = \bM^a \quad \text{where} \quad \bDiff \bM^a = 0\,.
\ee
To make further progress we will assume $\tau^2 <0$, so that we can call $\bDiff \tau^a=\bbe^a$ and have the coframe field at hand.
Then, since $\epsilon^{ad}{}_{bc}\+ \bR^c{}_d=2i\+ \bR^{a}{}_b$, we can write
\ba
 -2\bM^a & = & i\epsilon^{ac}{}_{bd}\+ \bR^b{}_c\wedge\bbe^d =  i\lp \+ R^a{}_{b}-\frac{1}{2}\delta^a_b\+ R\rp \ast\bbe^b\,. \label{giulini}
\ea
We have recovered the Einstein field equations for the self-dual curvature, sourced by a yet unknown 3-form $\bM^a$. 

It remains to show that this source term behaves as idealised dust.
By combining (\ref{solution}) with the (\ref{efe2}) we see that $\prescript{-}{}( \tau^{[a}\bM^{b]})=0$. At this point, we can pick the simplifying gauge $\tau^a=\tau\delta^a_0$, wherein it becomes apparent that the spatial 3-forms $\bM^I=0$ vanish. By construction (\ref{solution}) we have $\bDiff \bM^a =0$, which yields two further constraints, $\bomega^I{}_0\wedge\bM^0=0$ and $\bdiff \bM^0=0$. 
The former implies that $\bM^0$ is a spatial 3-form, $\bM^0=(i\rho/2) \ast\bbe^0$ for some function $\rho$, and the latter implies that this function $\rho$ dilutes with the spatial volume. Thus $\rho$ indeed  effectively describes the energy density of dust.     

Though the derivation was particularly transparent with the gauge choice $\tau^a=\tau\delta^a_0$, the conclusion naturally holds in any other gauge. We also have checked that coupling matter with (\ref{action})
would not change the form of $\bM^a$. 

\section{Cosmological constant}

Simply adding a constant $\Lambda$  {\it ad hoc} would not be compatible with first principles, but require the extension of the Lorentz to the de Sitter gauge group \cite{Koivisto:2021ofz}. Instead, we now more frugally supplement the action (\ref{action}) with two terms,
\be
\label{action2}
I = I_G - \frac{1}{24}\int \Lambda\epsilon_{abcd}\bDiff\tau^a\wedge\bDiff\tau^b\wedge\bDiff\tau^c\wedge\bDiff\tau^d - \bDiff\Lambda\wedge\bB\,.
\ee
Had we added only the first new term, the variation with respect to the $\Lambda$ would prohibit a viable spacetime by imposing that $\sqrt{-g}=0$. Therefore we also had to include the second new term, such that it rather imposes the constancy of $\Lambda$ with the 3-form Lagrange multiplier $\bB$. It is not difficult to see that the two new equations of motion are
\begin{subequations}
\label{efe}
\ba
\bdiff\Lambda & = & 0\,,  \label{efe3} \\
\bdiff \bB & = &  \frac{1}{24}\int \epsilon_{abcd}\bDiff\tau^a\wedge\bDiff\tau^b\wedge\bDiff\tau^c\wedge\bDiff\tau^d\,, \label{efe4}
\ea
\end{subequations}
and that the field equation (\ref{giulini}) is only modified by the addition of the appropriate $\Lambda$-term. Whilst (\ref{efe3}) ensures the constancy of $\Lambda$, the consequence of 
(\ref{efe4}) is unimodularity. In the broken phase it becomes $\partial_\mu \ast B^\mu = \sqrt{-g}$, and since the vector density $\ast B^\mu$ is not fixed, 
we are free to set $\sqrt{-g}=1$. --We note that the action (\ref{action2}) is nothing but the polynomial realisation of the well-known unimodular method \cite{Henneaux:1989zc} now embedded into the minimal Lorentz gauge theory (\ref{action}). 

The new field $\bB$ is a measure of global time. Consider a spacetime volume $V$ bounded by a hypersurface $\partial V$. Now, due to Stokes theorem,
\be
\oint_{\partial V} \bB = \int_{V} \sqrt{-g}\bdiff^4 x = \text{vol}{(V)}\,. \label{volume}
\ee
The spontaneous breaking of unimodular invariance, given by the normalisation chosen for $\partial_\mu \ast B^\mu$, determines the unit of global time. In the broken phase, we could
identify the {\it Kairon} scalar field $\sigma^a = \bbe^a\cdot\ast\bB$ in the fundamental representation of the Lorentz group on the same footing as the scalar field $\tau^a$.
Such a dual pair of fields arises from the embedding of the Lorentz gauge theory (\ref{action}) into the conformal gauge theory \cite{Koivisto:2019ejt}, and the intuition from previous studies 
\cite{Zlosnik:2016fit} also suggests the natural gauge choice $\bDiff\tau^a \sim \bDiff \sigma^a$. 

We have already reached the conclusion of this essay: the similar magnitude of the observed energy densities due to the $\Lambda$ and due to the CDM  \cite{Aghanim:2018eyx} could be the result of their common origin in the conformal geometry of the observer space. 

\section{Conclusion}

Rather than unknown particles or modified gravity, the dark sector of the universe could be the manifestation of a spontaneous symmetry breaking that underpins the emergence of a metric spacetime. 
In a rigorous derivation of the Einstein's field equations from a more fundamental, pre-geometric theory, both the $\Lambda$ and the CDM appear as integration constants.
Though spontaneous symmetry breaking has been considered as the origin for the difference between time and space \cite{Wetterich:2004za,Wetterich:2021hru}, similar results to ours have not, to our knowledge, been arrived at in less minimalistic settings.  

Of the 12 real components of the complexified Lorentz connection $\bomega^a{}_b$, the 6 self-dual pieces $\+\bomega^a{}_b$ account for the spin connection as usual, whereas 3 (the boosts $\prescript{-}{}\bomega^I{}_0$ in the $\tau^a=\tau\delta^a_0$ gauge) give rise to the spatial triad through $\bbe^a = \bDiff \tau^a$ in the broken phase. It remains to be seen whether the 3 remaining anti-self-dual rotations $\prescript{-}{}\bomega^I{}_J$ could be related to the $SU(2)_L$ connection in the particle sector, and whether the scalars $\tau^a$ and $\sigma^a$ could be related to the Higgs field. Another speculation is that our formulation might provide an improved starting point for loop quantum gravity that is currently suffering from a ``covariance crisis'' \cite{Bojowald:2021isp}.

To conclude, we propose a theory behind the two main parameters of the standard $\Lambda$CDM model of cosmology \cite{Aghanim:2018eyx}. 
\begin{itemize}
\item 1$^{\text{st}}$ $\Lambda$ problem, the sensitivity of gravity to vacuum energy, is resolved \cite{Weinberg:1988cp}.  
\item 2$^{\text{nd}}$ $\Lambda$ problem, the observed value of $\Lambda$, is related to the age of the Universe\footnote{The question of why the $\rho$ is small seems like the same question as why the universe is so old. The association of large numbers in physics with the age of the universe goes back, via Dirac, to Weyl. Recently, it was pointed out that Dirac's large number hypothesis might be realised in a model with two ``dilatons'' \cite{Koivisto:2021ofz}. However, here we only discussed the cosmological numbers related to the $\Lambda$ and the CDM.}.  
\item 3$^{\text{rd}}$ $\Lambda$ problem, the coincidence that $m^2_P\Lambda \sim \rho$, has a rationale in their dual origin.   
\end{itemize}
In particular, from the construction of the 3-form $\bM^a$ we have that $\partial\tau \sim \sqrt{m_P/\rho}$, and from (\ref{volume}) we read $\partial_\mu \sigma ^\mu \sim 1/\sqrt{\Lambda}$.    
The duality $\tau \sim \sigma$ could explain the cosmic coincidence.  

\acknowledgements{This work was supported by the Estonian Research Council grants PRG356
``Gauge Gravity'' and MOBTT86, and by the European
Regional Development Fund CoE program TK133 ``The
Dark Side of the Universe''.}

\bibliography{Krefs}

\end{document}